\documentclass[12pt,letterpaper]{article}

\usepackage{datetime}
\usepackage{amsmath,amssymb,array,calc,rotating,epsfig,psfrag, amscd}
\usepackage{color}
\usepackage{verbatim}
\usepackage[colorlinks=false,
    linktoc=all, 
    pdfstartview=FitV,
    bookmarksopen=true]{hyperref}
\usepackage[left=2cm,top=1cm,right=3cm,nohead]{geometry}
\usepackage[english]{babel}

\newcommand{\bea}{\begin{eqnarray}}
\newcommand{\eea}{\end{eqnarray}}
\newcommand{\be}{\begin{equation}}
\newcommand{\ee}{\end{equation}}

\definecolor{cardinal}{rgb}{0.6,0,0}
\definecolor{darkgreen}{rgb}{0,0.5,0}
\definecolor{golden}{rgb}{0.92, 0.7, 0}
\definecolor{midnight}{rgb}{0, 0, 0.5}
\definecolor{darkblue}{rgb}{0.2, 0, 0.8}

\newcommand{\beq}{\begin{equation}\begin{aligned}}
\newcommand{\eeq}{\end{aligned}\end{equation}}

\numberwithin{equation}{section}

\begin{document}
\begin{flushright}
\small
UUITP-02/15
\normalsize
\end{flushright}

\thispagestyle{empty}

\begin{center}
\baselineskip=13pt {\LARGE \bf{Perturbative decay of anti-branes in flux backgrounds due to space time instabilities}}
 \vskip1.5cm 
Ulf Danielsson \\
\vskip0.5cm
 \textit{Institutionen f{\"o}r fysik och astronomi,\\ Uppsala Universitet, Uppsala, Sweden}\\
\href{mailto:ulf.danielsson@physics.uu.se}{ulf.danielsson @ physics.uu.se}

\vskip3.5cm
\end{center}

\begin{abstract}
\noindent
In this paper we suggest a new source of perturbative decay of the KPV-state, which might have consequences for the viability of the KKLT-construction. The results do not rely on any direct enhancement of the decay due to flux accumulating on the anti-brane in transverse space. Instead, we note that the system can lower its energy through a sequence of NS5-configurations all the way to the true vacuum, without encountering a barrier, if we allow for clumping of screened charge in space time. The clumping can possibly be a parallel to the Gregory-Laflamme instability of black branes. The results are obtained at large $p$, but for $p/M$ arbitrarily small. It is furthermore argued that the results extend to cases of few or single anti-branes where quantization becomes important. We believe that it is important to investigate this possible effect further to judge whether there are any fatal consequences.

\end{abstract}

\clearpage

According to KPV, \cite{Kachru:2002gs}, there is a way for anti-branes near the tip of a flux supported Klebanov-Strassler background, \cite{Klebanov:2000hb}, to decay, if the anti-branes polarize into an NS5-brane that moves across an internal $S^3$. There is a competition between the cost in energy to expand the NS5 across the $S^3$, and the force due to the background $H$-field. The result is a barrier that requires tunneling to get through.

In several works it has been observed that the $H$-field blows up near the anti-brane, see, e.g.,  \cite{Bena:2009xk, McGuirk:2009xx, Bena:2010gs, Bena:2011hz, Bena:2011wh, Blaback:2011nz,  Blaback:2011pn, Giecold:2011gw, Massai:2011vi, Blaback:2012nf, Massai:2012jn, Bena:2012bk, Gautason:2013zw, Blaback:2013hqa, Cottrell:2013asa, Dymarsky:2013tna,  Giecold:2013pza,  Blaback:2014tfa, Danielsson:2014yga, Bena:2014jaa}. There has been a long standing debate about whether this is a physical effect or not, but the conclusion seems to be that the effect is real. Physically, it can be understood as the accumulation of charge carried by the fluxes $H$ and $F_3$ towards the anti-brane. 
 
A remaining question is whether this pile up of flux will affect the stability of the system. It has been argued that the increase in the $H$ field will increase the force on the NS5 brane, and lower the barrier against annihilation. In this paper we will not take such effects into account, but instead focus on another possible instability indirectly associated with the accumulation of flux. The fate of this possible instability will be of importance for the KKLT-construction of dS-space, \cite{Kachru:2003aw}.

A clue to what might happen is that the amount of accumulated flux on the anti-brane is just right to completely screen the anti-brane when viewed from a distance in transverse space, \cite{Michel:2014lva}. As a consequence the system is, at least in some respects, very similar to a neutral black-brane. The properties of charged and neutral black branes in flux-backgrounds have been studied in, e.g., \cite{Aharony:2007vg, Buchel:2010wp, Bena:2012ek, Bena:2013hr, Hartnett:2015oda}, hinting at many interesting properties. Even though they are neutral, one can still speculate about their actual composition at a microscopic level. How much of the antibrane charge has actually been able to annihilate against the charged flux? It is interesting to compare with the microscopic picture of standard black branes in empty space proposed in \cite{Danielsson:2001xe}. There, branes and anti-branes are put on top of each other, but they nevertheless do not annihilate instantaneously. One can understand this from a thermodynamical point of view, where the temperature of the system provides the tachyonic open strings between the branes and anti-branes with thermal masses. The energy of the black brane is carried by the tension of the composite branes, and by the open string gas on top of them. The exact proportion is determined thermodynamically. Eventually the system decays through Hawking radiation as the open strings are converted into closed strings.. 

However, there is another interesting clue to the physics of the KPV-state that is suggested by the parallel with black branes. As discovered in \cite{Gregory:1993vy,Gregory:1994bj} by Gregory and Laflamme, a sufficiently extended black brane is unstable against clumping. For instance, it is suggested that the horizon of a black string develop irregularities, which leads to the black string transforming into a row of black holes. If the black string is carrying charge the end result is a string of pearls, where the pearls are black holes. This can be argued by considering perturbations of the metric, or by using a thermodynamical argument. In \cite{Danielsson:2001xe} the Gregory-Laflamme transition was used to derive properties of black holes in ten and eleven dimensions, through the use of unstable black D3-branes, black M2- and black M5-branes, respectively. Could there be a similar instability in the KPV-state? Instead of pursuing the parallel above further, we will try to answer this question using the microscopic description of the system in terms of NS5-branes.

The action we will use for the NS5-brane is given by  \cite{Kachru:2002gs},
\be
\frac{\mu_5}{g^2} \int d^6 \xi  \sqrt{-{\rm det} (G_\parallel){\rm det} (G_\perp+2\pi g {\mathcal F})} + \mu_5 \int B_6 ,
\ee
where $G_\perp$ refers to two directions of the NS5 that wraps an $S^2$ on the $S^3$, while $G_\parallel$ refers to the other four directions of the NS5. These will stretch over space time such that the remaining coordinate on $S^3$, the angle $\psi$, in general will be a function of the space time coordinates. These will be time, two directions parallel, and one transverse, to what eventually will be a domain wall between the false and the true vacuum.  We have that
\be \label{Fshift}
2\pi {\mathcal F}_2= 2\pi F_2 -C_2 ,
\ee
where $F_2$ is the field strength of the world volume gauge field, and $C_2$ the gauge potential for the space time field $F_3$. It is ${\mathcal F}_2$ that tells us how much D3-brane charge is effectively carried by the NS5. As the NS5 moves across the internal $S^3$ from one pole to the other, the initial charge $-p$ of the anti-branes, is converted into a net brane-charge of $M-p$. This is the number of branes that can nucleate when the NS5 reaches the other pole of the $S^3$, and the transition to the true vacuum is complete. The radius of the $S^3$ is given by $R\sim \sqrt{gM}$.

In \cite{Kachru:2002gs} it is shown that the potential for a static configuration, with a constant angle $\psi$, is given by
\be \label{VKPV}
V\sim \frac{1}{\pi} \sqrt{b_{0}%
^{4}\sin^{4}\psi+ \bigl(\pi\frac{p}{M}-\psi+\tfrac{1}{2}\sin(2\psi)
\bigr)^{2}} - \tfrac{1}{2\pi}(2\psi-\sin(2\psi))+\frac{p}{M}\, ,
\ee
where $b_0^2\simeq 0.93266$. See \cite{Herzog:2001xk} for more details on the geometry. Here, and in the rest of the paper, we are expressing the potential in units where the uplift is close to $\frac{2p}{M}$. A detailed study of the potential shows that there is a barrier for all values of $p$ smaller than a critical value given by $p \sim 0.08 M$, where $M$ is the flux number for $F_3$. As a consequence, any decay is non-perturbative for values smaller than this. In particular, for values relevant for KKLT, e.g. $p$ of order one, there is a very heavy suppression of the decay. The pile up of flux towards the anti-brane (or charged NS5) mentioned above may change this in some regimes, but, for simplicity, we will omit this effect in the present work. It has been argued in \cite{Michel:2014lva} that the effect will be negligible if $p$ is small. \footnote{I thank Joe Polchinski for explanations of the work in \cite{Michel:2014lva}.}

We will generalize the above to configurations where $\psi$ is no longer a constant as a function of space time. Furthermore, and this we believe is new, we will also allow for  the density of screened flux to vary along a direction in space, and hence on the NS5. We parametrize the density using $p$, and allow $p$ to become higher in some regions of space, and lower in others. We assume that the integral of $p$ over space remains the same. Note that $p$ always refers to the initial anti-brane charge not counting the contribution coming from $C_2$. Our idea is to compare the energy for different configurations of the NS5 brane, and investigate whether it is possible to complete the transition without encountering a barrier. 

Ignoring polarization, the energy as given by (\ref{VKPV}) is simply proportional to $p$, and the integrated value does not change. When we allow the NS5 brane to polarize, and relax to the value of $\psi$ that locally minimizes the potential, there will be a slight lowering of the energy. As we will see the effect of lowering the energy is larger for larger values of $p$. This is connected to the fact that the barrier even disappears for $p$ large enough. As a consequence, the energy will depend on the details of the clumping. In fact, our argument suggests that {\it configurations with heavily peaked distributions of screened charge will have lower energy}. Below we will verify this numerically.

To demonstrate our idea we will consider variations in only one direction.  This direction can be viewed as being transverse to a domain wall that ultimately will separate the false vacuum from the true one. As we move across such a domain wall, the NS5 will move across the $S^3$ from a position near one pole all the way to the other pole. These kind of configurations were discussed in detail in \cite{Danielsson:2014yga}, where it was also explained how the effective tension of the domain wall, from the 4D perspective, is related to the strength of the barrier.

Let us now examine this effect from the point of view of the NS5-brane, and make use of the KPV-potential to show that there are paths in configuration space such that the system can lower its energy monotonically, while the NS5 goes completely over the $S^3$ in a limited region, thus initiating the decay. To do this we evaluate the energy of some interesting configurations. For simplicity, we will assume that a transverse distance $L$ is divided into two regions such that $L=L_1+L_2$. Initially, we have a constant screened charge $p<<M$ across the whole distance, which will dissolve into the NS5 that in turn polarizes slightly. We will then examine what happens to the energy of the system if we let the system clump such that $p_1<p$ across $L_1$, and $p_2>p$ across $L_2$, with the total charge the same, i.e., $p_1L_1+p_2L_2=pL$. We will then examine different profiles for the $NS5$ and compare their energies. 

Figure 1 shows the case with constant $p$ together with the polarized NS5. The energy of the configuration is given, in terms of the KPV-potential, by
\be
E_0 = L V(p,\psi _{min}),
\ee
where $\psi _{min}$ is the value of $\psi$ that minimizes the value of the potential. In the figure, the small circle represents the $S^2$ positioned at angle $\psi$ on the $S^3$. The linear direction is transverse to the domain wall that eventually will appear. We have also two dimensions in the plane of the domain wall that we have suppressed. 

\begin{figure} \center
\includegraphics[height=75mm]{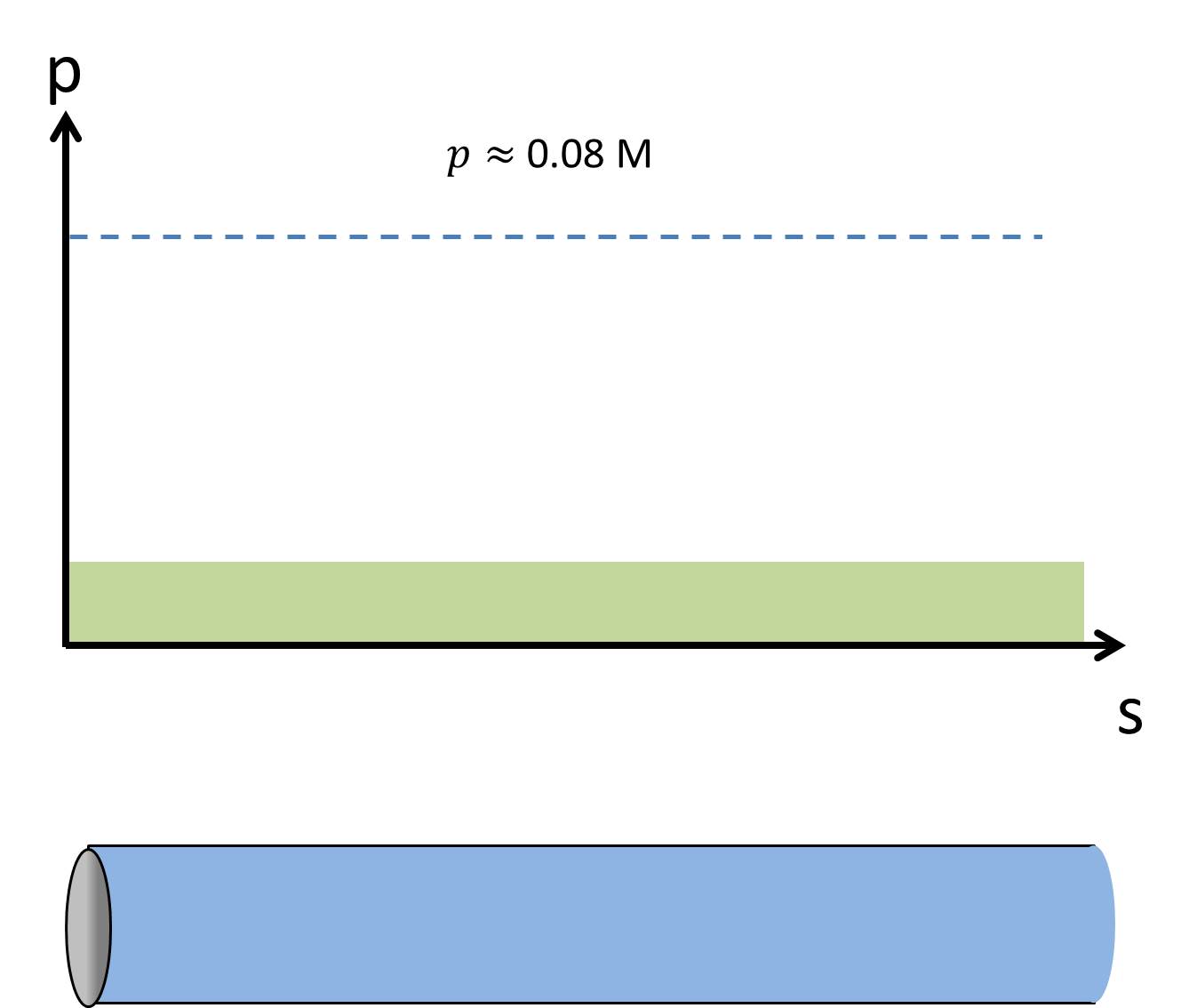} 
\caption{\emph{The constant configuration we start with, together with the slightly polarized NS5.} }%
\end{figure} 

\begin{figure} \center
\includegraphics[height=100mm]{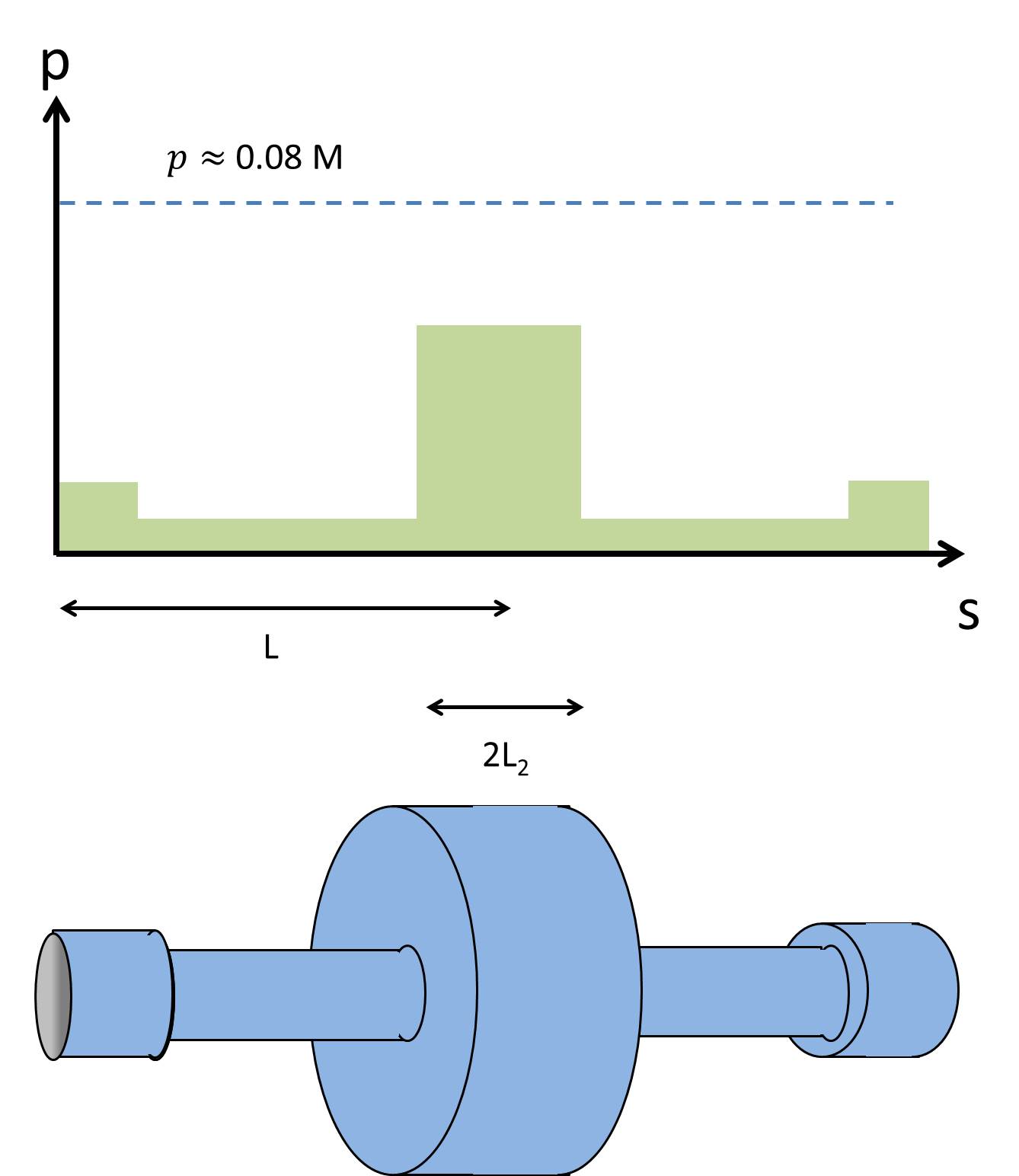} 
\caption{\emph{A peaked configuration of screened charge, together with the corresponding configuration of the NS5 that does not lead to a transition to the true vacuum.} }%
\end{figure} 

Figure 2 shows what we have in mind to describe the clumping. Here, we have two copies of the configuration to facilitate the interpretation of the process as bubble nucleation of the true vacuum. We assume that the NS5 is positioned in such a way that, given $p_i$ and $L_i$, it minimizes the energies separately in the two regions.  We see in the figure how it reacts to the differing values of $p$. Note that, in this example, the NS5 reaches a maximum value of $\psi$ in the middle, and goes back down again ending near the same pole on the $S^3$ where it started. This particular configuration, therefore, does not represent a transition.

As we will see, and this is the main result of the paper, the configuration in figure 1 is a solution of the equations of motion, but an unstable one. This means that we can {\rm lower} the energy by considering violations of the equations of motion for the position $\psi$ of the NS5, as a function of space, and the field strength $F_2$ on the NS5. The violation of the equations of motion for such a static configuration acts like a force that will drive the system to even lower energies.

To calculate the energy in cases where we have a nontrivial dependence on the shape of the NS5 brane as a function of space, $\psi = \psi (s)$,  as well as a non-trivial distribution of charge, $p=p(s)$, we must re-examine the expression for the potential. We let the three world volume coordinates on NS5 that lie along the $S^3$ be $(\xi , \theta , \phi )$. For the angles $\theta$ and $\phi$ we do not make a distinction between the coordinates on the $S^3$, and the corresponding coordinates on the NS5. In the remaining direction, however, the coordinate $\psi$ on the $S^3$, and the coordinate $s$ in space, are nontrivial functions of $\xi$. 

The charge carried by the NS5 is due to the components ${\mathcal F}_{\theta \phi}$ of the shifted two-form gauge field in (\ref{Fshift}). We have
\be \label{F2int}
2\pi \int_{S^2} F_2 = 4\pi ^2 p ,
\ee
and want this integral to vary as we move in $\psi$ over the $S^3$. At first sight, this seems to violate charge conservation from the point of view of the NS5. After all, $F_2$ can be viewed as being sourced by a magnetic monopole charge $-p$ sitting at the pole, and the integral in (\ref{F2int}) should be independent of where the $S^2$ is positioned. This is true if only the components $F_{\theta \phi}$ are turned on, but need not be the whole story. We see this from solving the Bianchi identities for the magnetic field using a gauge potential such as, e.g., $A_\phi = p \cos \theta $. If we let $p$ be a function of $\psi$ (violating the static equations of motion in accordance with our strategy), we will generate new terms of the form $F_{\psi \phi}$ (and in general $F_{\psi \theta})$, which will cost energy. These will act like new sources that can screen and anti-screen the charge at the pole.  There is only one consistency condition that needs to be fulfilled, we impose $\int _{S^3 /\{ \rm{NP,SP}\} } dF_2  =0$, outside the singular sources at the poles. This is automatic since $p$ is a single valued function that will integrate to its original value. In other words, the screening charges mimicked by $F_{\psi \theta}$ and $F_{\psi \phi}$ sum up to zero.

While we have argued that a varying $p$ is consistent with charge conservation, and the Bianchi identity, there will still be a cost in energy to make these deformations. We now turn to an estimation of this.  If we evaluate the NS5 action in the presence of these new terms, the piece responsible for the tension of the NS5 will be proportional to
\be
\int d\xi \sqrt{ G_{\xi \xi} (G_{\theta \theta}G_{\phi \phi}+{\mathcal F}_{\theta \phi}^2)+G_{\theta \theta}{\mathcal F}_{\phi \xi}^2+G_{\phi \phi}{\mathcal F}_{\theta \xi}^2} .
\ee
Here we have been careful in expressing everything in terms of the world volume coordinate $\xi$ on the NS5, with $\psi$ a function of $\xi$. From this we conclude that the induced metric on the NS5 is given by
\be
 G_{\xi \xi}=\left(\frac{d\psi}{d\xi}\right)^2 G_{\psi \psi} +\left(\frac{ds}{d\xi}\right)^2 G_{ss} =\left(\frac{ds}{d\xi}\right)^2\left(G_{ss} +\left(\frac{d\psi}{ds}\right)^2 G_{\psi \psi}\right),
\ee
while the new components of ${\mathcal F}_2$ are given by
\be
{\mathcal F}_{\xi \theta} = \frac{d\psi}{d\xi} F_{\psi \theta} = \frac{ds}{d\xi}\frac{d\psi}{ds} F_{\psi \theta}.
\ee
This gives
\be
\int ds \sqrt{ \left(G_{ss} +\left(\frac{d\psi}{ds}\right)^2 G_{\psi \psi}\right) \left(G_{\theta \theta}G_{\phi \phi}+{\mathcal F}_{\theta \phi}^2\right)+G_{\theta \theta}\left(\frac{d\psi}{ds} F_{\psi \theta}\right)^2+G_{\phi \phi}\left(\frac{d\psi}{ds} F_{\psi \phi}\right)^2}
\ee
When $\psi$, and $p$, are constant we can simply drop all terms with $\frac{d\psi}{ds}$, and all terms with ${\mathcal F}_2$ except the ${\mathcal F}_{\theta \phi}$. We are then back at ({\ref{VKPV}). However, at the points where $p$ makes a jump these terms will contribute. In fact, in the limit of a sharp jump, the expression will be dominated by 
\be
\int ds \frac{d\psi}{ds} \sqrt{G_{\psi \psi} (G_{\theta \theta}G_{\phi \phi}+{\mathcal F}_{\theta \phi}^2)+G_{\theta \theta} F_{\psi \theta}^2 +G_{\phi \phi} F_{\psi \phi}^2},
\ee
which is the cost of the jump. The integral over the sharp jump picks up a factor that tells us how far $\psi$ moves during the jump, which in general will be of order $1$. This multiplies an expression that will be of order $R$, and not scale with $L$. Note that $F_{\psi \theta}$ and $F_{\psi \phi}$, are finite when expressed using an index $\psi$ rather than $\xi$. In units where the uplift is of order $\frac{p}{M}$ the maximum is at order $1$. So, we find a contribution of order $R$ from the discrete jump, while, as we will see, the gain in energy will scale like $L$. Hence, if we take $L/R$ sufficiently large the contributions from the steps can be neglected. This is a crucial part of the argument, which we will come back to later.

The dominating contributions (when $L>>R$) then gives
\be
E_1 = L_1 V(p_1, \psi _{1, min})+L_1 V(p_1, \psi _{2, min}) ,
\ee
to be compared with the value $E_0$ for the false vacuum. Our main result is that we can construct a continuous path, starting at $E_0$,  where the energy $E_1$ along the path monotonically decreases all the way until the transition is completed.

We will now consider a numerical example. We choose $p=1$ and $M=1000$, which is well within the regime where, according to KPV, the system is meta-stable with a huge barrier, and cannot decay classically. Figure 3 shows the results, and we clearly see the promised decrease in the energy  as the clumping proceeds.

\begin{figure} \center
\includegraphics[height=50mm]{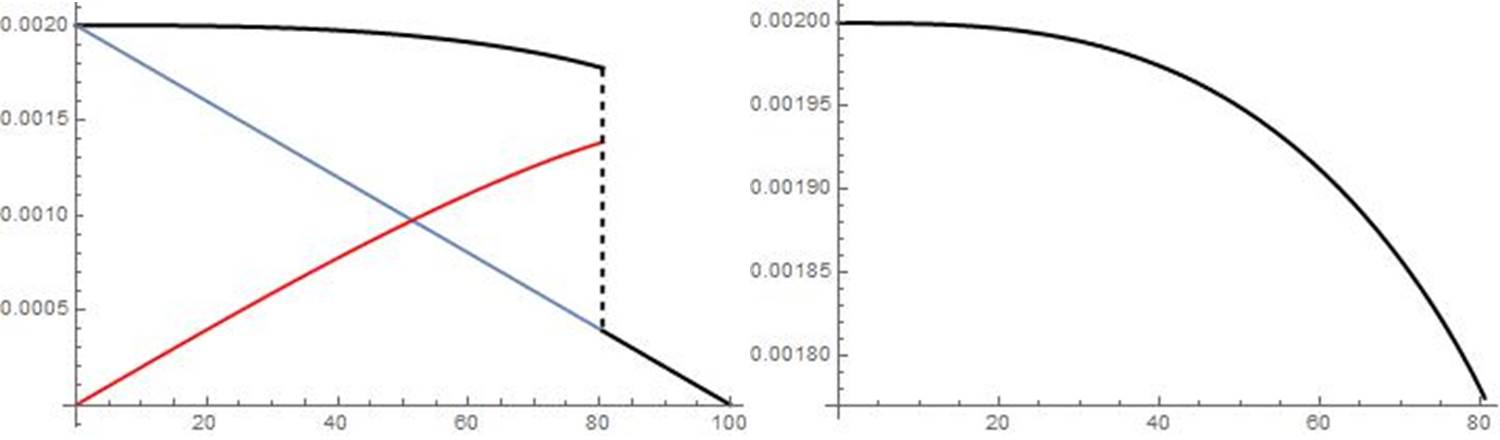} 
\caption{\emph{A plot of the total energy as a function of $p_2$ is shown in black, demonstrating the instability of the configuration in figure 1. The contribution from the piece of space with $p=p_1$ is shown in red, while the contribution from the piece with $p=p_2$ is shown in blue. Note how the increase in energy in the first region is compensated by a slightly larger decrease in the second region. } }%
\end{figure}

To further illustrate what is happening, table 1 compares the energy for a few selected configurations where $L_1=0.99$ and $L_2=0.01$ are held fixed, while varying $p_1$ and $p_2$. The corresponding potentials for these cases are shown in figures 4-7. In the final example the contribution to the energy from the piece $L_2$ vanishes, since the barrier is now gone, and the NS5 has collapsed at the other pole. This is simply because the local ratio $p/M$ is enhanced above the critical value of $\sim 0.08$. In figure 8 we see the configuration of screened charge and the NS5, respectively.  From this point on the system can lower its energy simply by letting the bubble of true vacuum expand. 

\begin{table}[t!]
\renewcommand{\arraystretch}{1.35}
\begin{center}
\begin{tabular}{|c|c|c|}
\hline
$p_1$ & $p_2$ & Energy \\
\hline
\hline
1 & 1 & 0.00199996\\
0.909 & 10 &  0.00199957 \\
0.404 & 60 & 0.00191186 \\
0.0101 & 99 & 0.00002\\

\hline
\end{tabular}
\end{center}
\caption{The total energy for some different configurations of screened charge.}
\end{table}

\begin{figure} \center
\includegraphics[height=50mm]{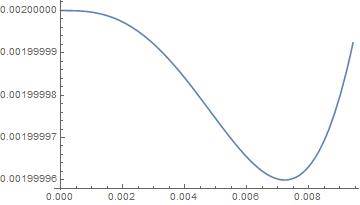} 
\caption{\emph{The minimum of the potential for our example with $p_1=p_2=1$.} }%
\end{figure}

\begin{figure} \center
\includegraphics[height=50mm]{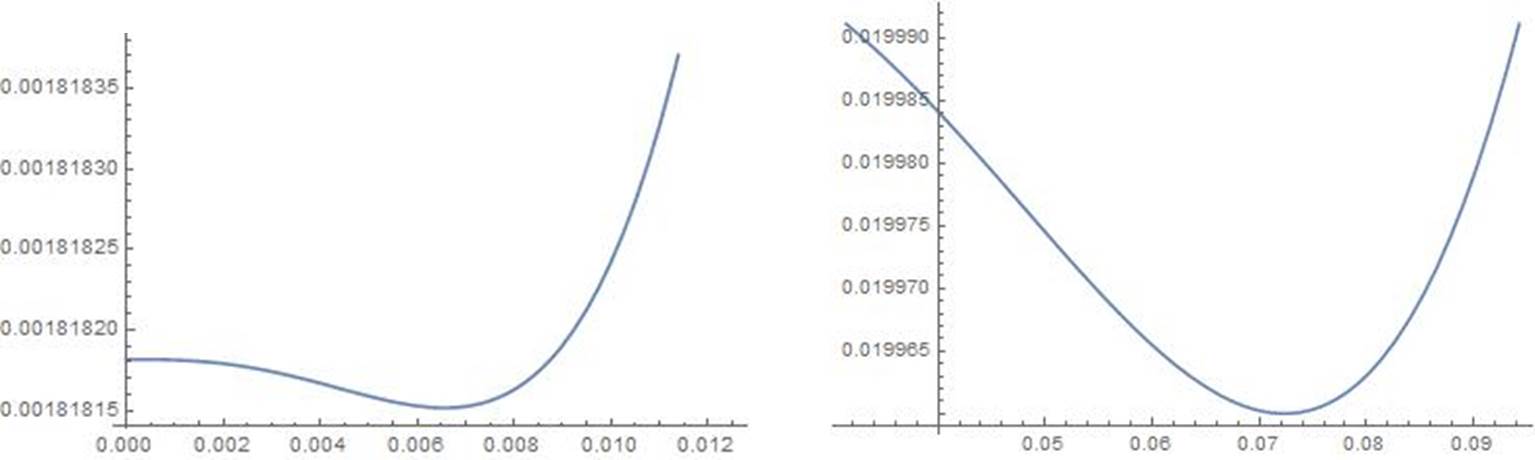} 
\caption{\emph{The minimum of the potential for our example with $p_2=10$ is shown on the right. The minimum for the corresponding value for $p_1$ is shown on the left.} }%
\end{figure}

\begin{figure} \center
\includegraphics[height=50mm]{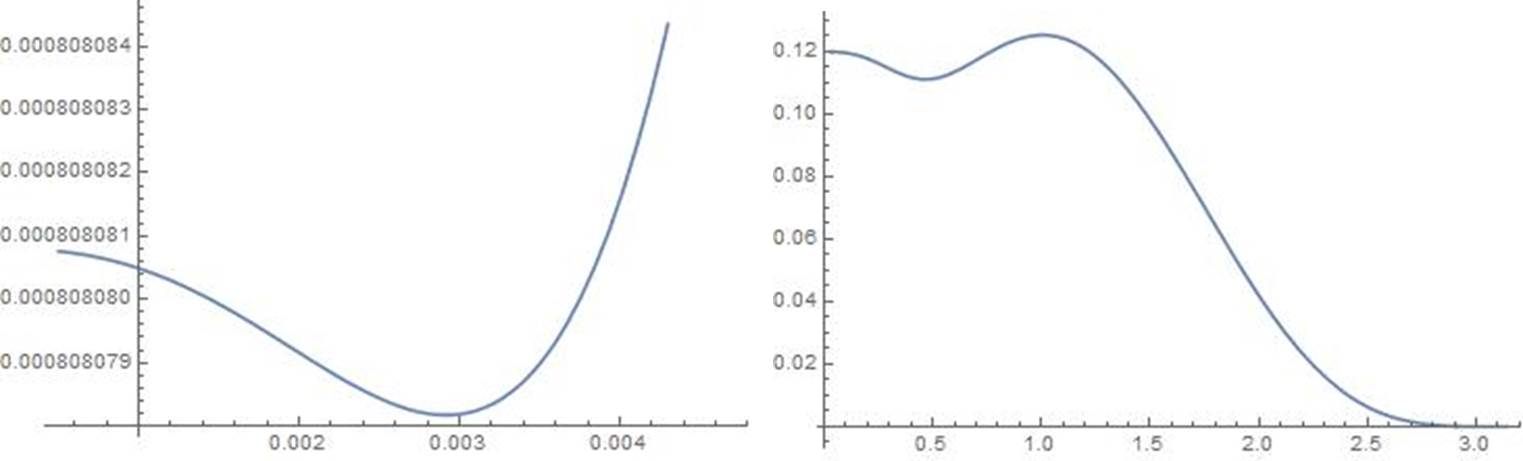} 
\caption{\emph{The minimum of the potential for our example with $p_2=60$ is shown on the right. The minimum for the corresponding value for $p_1$ is shown on the left.} }%
\end{figure}

\begin{figure} \center
\includegraphics[height=50mm]{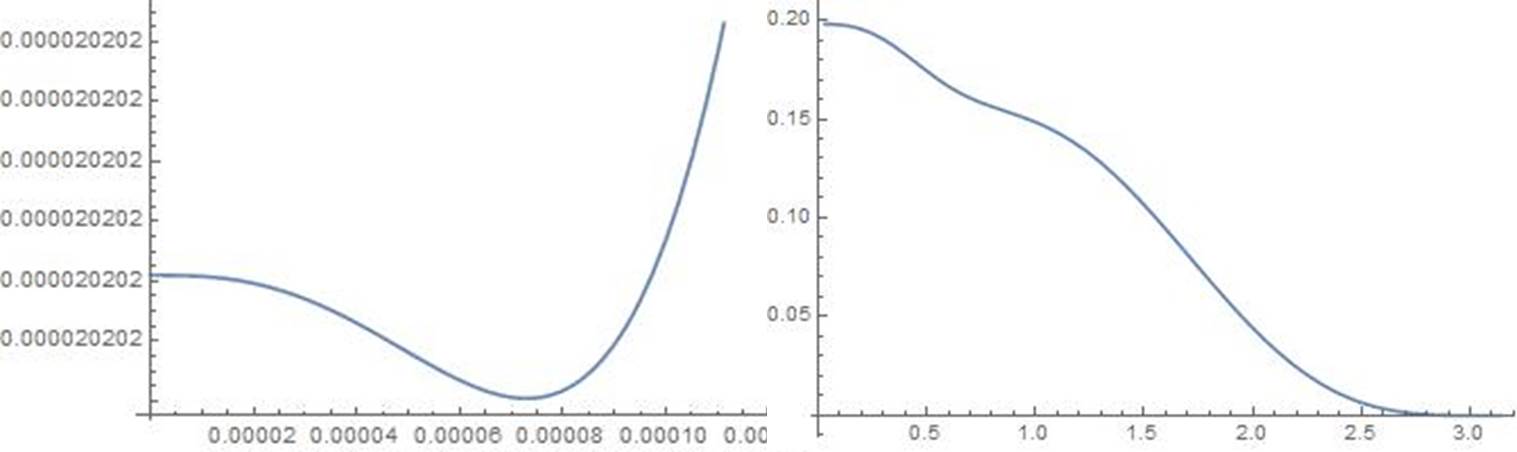} 
\caption{\emph{The potential, without a barrier, for our example with $p_2=99$ is shown on the right. The minimum for the corresponding value for $p_1$ is shown on the left.} }%
\end{figure} 

We are now ready to make an estimate of how large $L$ needs to be. We have a cost in energy from the jumps of order $R$,
which should be compared with the reduction in the energy of the configuration that just made the transition, $ \sim L_2\frac{p}{M}$, and hence we need $L_2 \gg \frac{MR}{p}$. On the other hand, we need roughly $L/L_2 \sim 10^{-1} \frac{M}{p}$ to remove the barrier, and therefore $L \gg 10^{-1} \left( \frac{M}{p} \right) ^2 R$. In our particular example, this translates into $L \gg 10^{5} R$. The inequality is to make sure that the terms we neglected are small compared to the ones we kept, but one should not that the instability shows up already when the inequality is saturated.

\begin{figure} \center
\includegraphics[height=100mm]{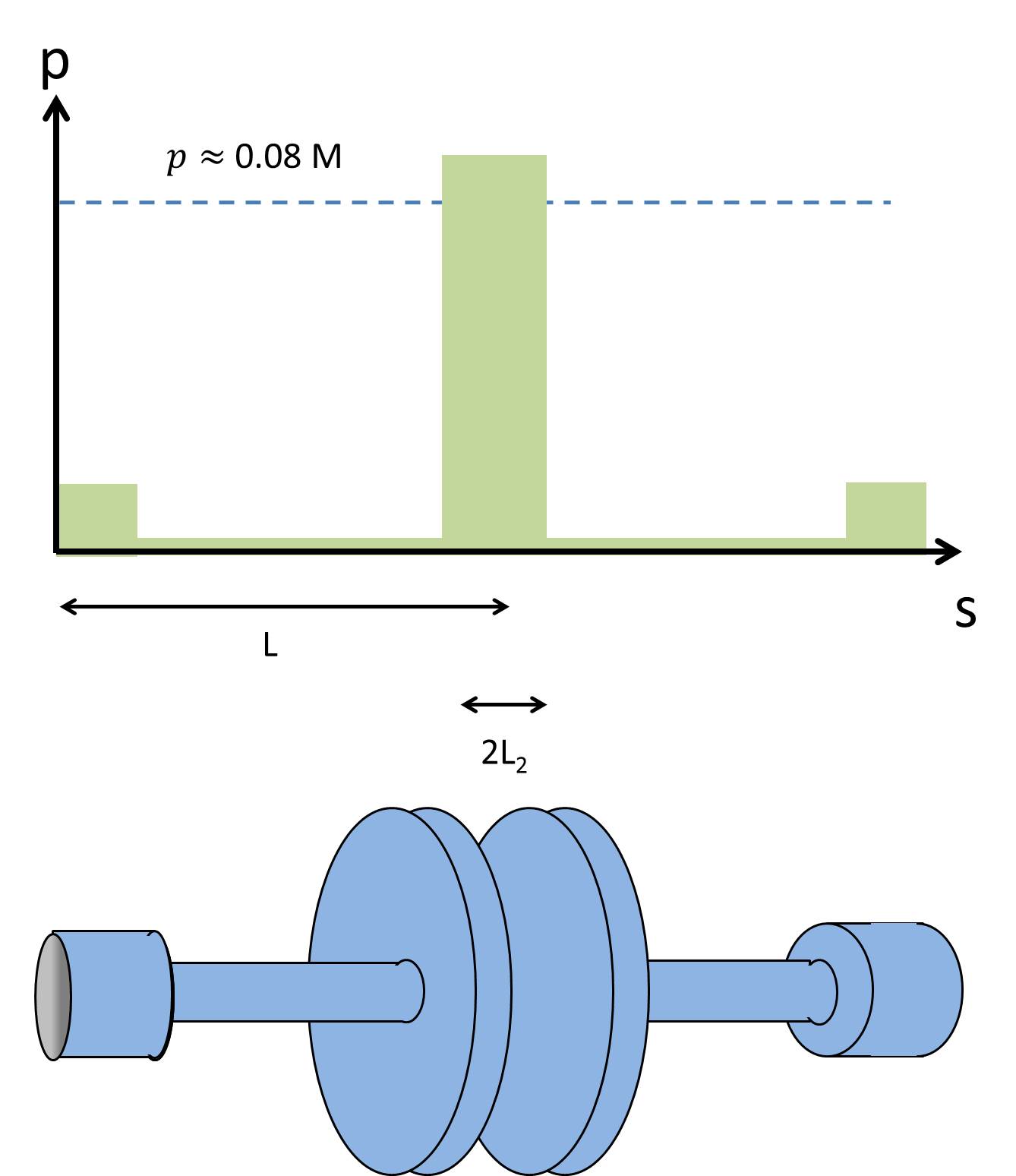} 
\caption{\emph{A peaked configuration of screened charge that leads to a transition to the true vacuum.}}%
\end{figure}

What role does the pile up of flux responsible for the screening actually play? In our calculation it does not play a direct role, but is just an indirect, even though unavoidable, consequence of the presence of the background flux.  Are there corrections that we have not taken into account that could be due to the pile up? This has been the subject of some controversy. In principle the presence of the pile up could speed up the process, but \cite{Michel:2014lva} has argued that such an effect would be very small -- at least for small $p$.

From a space time point of view, at large distances, the presence of the screening makes it natural to interpret the process as a Gregory-Laflamme transition. The scale $L$ of the instability is much larger than $R$, and thus can be viewed as a property of a neutral fluid. The fluid contains components that are mutually non-BPS so that the gravitational force is not balanced. It is therefore natural to expect the system to gain energy when it clumps. If this effect is fully taken into to account the process we have discussed should proceed with even greater ease.

One concern with the KPV-approach, as well as with KKLT, is the validity of the calculations at small $p$, such as $p=1$, where quantization plays a role. In the numerical calculations of this paper we formally put $p=1$, and discovered a possible instability. To safely sit in the classical regime one should take $p$ large, and scale $M$  such that $p/M$ remains the same. The removal of the barrier depends on this ratio only, and we have seen that the $p \sim 0.08M$ is easily circumvented by using peaked profiles in space time. 

So, what about $p$ small, such as $p=1$? One possibility is that we in the quantum mechanical treatment need to consider discrete transitions. For instance, one could imagine that we, in our example above, directly make a transition from $p=1$ to a state with $p_1=0$ and $p_2=100$, with $L_2=0.01L$. Our analysis shows that even though the quantum system makes a direct jump, there is a classical path that connects these states. Alternatively, one can connect the two states by choosing $p_1=0$, and $p_2=1/L_2$, while adjusting $L_2$ in discrete jumps all the way up to $p_2=100$. It is easy to convince oneself that this is also a trajectory with monotonically decreasing energy. One can then ask whether the time it will take to make these quantum transitions will differ in a significant way from the time it takes for a classical transition.

It is instructive to compare with positronium. Classically, the electron and the positron can fall closer to each other, shifting from one orbit to another lowering their energy, while emitting radiation. In quantum mechanics, the transitions occur, in contrast, in discrete jumps. However, it is not difficult to see that the time it takes for the classical fall, together with the emission of radiation according to the Larmor formula, is of the same order as the average time it takes for the quantum transition. Hence, the quantization as such need not necessarily change the life time of a state in a significant way. An exception is, of course, ground states. The crucial result of our analysis is that there is a classical path without a barrier connecting the states. The importance of this fact remains even if we take the quantization of $p$ into account.

Our argument suggests that quantization need not be an issue, and that there is nothing special with $p=1$. Is there a loophole? There has been suggestions that, in contrast to $p>1$, the case $p=1$ does not polarize and behaves in a completely different way. Support for this point of view comes from interpreting the work of \cite{Myers:1999ps, Polchinski:2000uf} as requiring at least $p=2$ to give polarization. This is in contrast with the NS5-picture, which does not indicate such an unexpected difference.  A conservative interpretation would be that the somewhat unnatural discrepancy at $p=1$ is an artifact, and not solid enough to be the foundation of a stability argument. Nevertheless, this seems to be among the most crucial questions to sort out if the KKLT-construction is to be put on firm ground.
 
It is important to note that our result is robust with respect to several kinds of corrections to the potential. Terms with higher order derivatives would contribute to the kinks in the profile, but such corrections can be made sub-leading simply by choosing $L$ large enough. Corrections important when $p/M$ is small, will mostly affect the part of the profile where $p=p_1$.  Adopting profiles with $p_1=0$, and varying $L_2$, the importance of such corrections can be reduced. 

One should also stress that the process involves small energy differences, with length scales that are large compared to the string scale and $R$, the size of the $S^3$. Nevertheless, all of these scales are microscopic. It is natural to expect that the characteristic time scale of the process will be given by scales like this, and the uplift. This would still give a much shorter time than the Hubble time corresponding to the dark energy. The uplift given by the anti-brane, as calculated by KPV, is carefully balanced against the negative energy of a supersymmetric AdS-vacuum. Even a tiny perturbation in the KPV-energy can translate into a huge change in the effective cosmological constant. This puts further constraints on attempts to construct a realistic model despite the instability. 

Let us summarize what our claim is. Starting with a homogeneous distribution of screened charge, the system can continuously lower its energy all the way to the true vacuum without encountering a barrier. Hence, we expect an instability towards spontaneous creation of bubbles of true vacuum. It seems important to investigate this possible effect further to see whether it puts constraints on the KKLT-construction. In particular it would be interesting to see whether there is a connection with the Gregory-Laflamme instability.

\section*{Acknowledgements}
I would like to thank Johan Bl{\aa}b{\"a}ck, Giuseppe Dibitetto, Daniel Junghans, Thomas Van Riet, Sergio Vargas, and Timm Wrase for important comments, and stimulating discussions.  The work is supported by the Swedish Research Council (VR).

\bibliography{BiblioThomas}
\bibliographystyle{utphys}

\end{document}